\documentclass[12pt,a4paper]{article}

\newcommand{\la}{\lambda}
\newcommand{\rx}{\mathrm{X}}
\newcommand{\de}{\Delta}
\newcommand{\pa}{\partial}
\newcommand{\del}{\delta}
\newcommand{\xa}{\xi_1}
\newcommand{\xb}{\xi_2}
\newcommand{\xc}{\xi_3}
\newcommand{\al}{\alpha}
\newcommand{\ms}{\mathcal{S}}
\newcommand{\mj}{\mathcal{J}}

\begin{document}

\begin{flushright}
{}
\end{flushright}
\vspace{1.8cm}

\begin{center}
 \textbf{\Large Three-Point Correlator of Heavy Vertex Operators \\
for Circular Winding Strings in $AdS_5 \times S^5$ }
\end{center}
\vspace{1.6cm}
\begin{center}
 Shijong Ryang
\end{center}

\begin{center}
\textit{Department of Physics \\ Kyoto Prefectural University of Medicine
\\ Taishogun, Kyoto 603-8334 Japan}  \par
\texttt{ryang@koto.kpu-m.ac.jp}
\end{center}
\vspace{2.8cm}
\begin{abstract}
We consider an extremal three-point correlator of three heavy
vertex operators for the circular winding string state
with one large spin and one windining number in $AdS_5$ and one
large spin and one winding number in $S^5$. We use a 
Schwarz-Christoffel transformation to compute semiclassically the
extremal three-point correlator on a stationary string trajectory 
which is mapped to the complex plane with three punctures. 
It becomes a 4d conformal invariant three-point correlator on the 
boundary. We discuss the marginality condition of vertex operator.
\end{abstract} 
\vspace{3cm}
\begin{flushleft}
April, 2012
\end{flushleft}

\newpage

In the AdS/CFT correspondence \cite{JM} a lot of fascinating results
have been found in the computation of the planar contribution to the 
conformal dimensions of non-BPS operators for any value of the coupling
constant, using integrability \cite{NB}.

Due to the AdS/CFT correspondence the correlation functions in the 
$\mathcal{N}=4$ super Yang-Mills theory (SYM) can be calculated both
at weak and at strong coupling. The three-point correlation functions
of BPS operators have been derived at strong coupling in the 
supergravity approximation. There have been various 
investigations that correlation functions of operators in the
$AdS_5 \times S^5$ string theory carrying large charges of order of string
tension $\sqrt{\la}/2\pi$ should be controled at large $\la$ by the
semiclassical string solutions \cite{GK,DB,DSY,AAT}.

The two-point correlators of heavy string states have been constructed
by using the vertex operator approach \cite{EB} and the wave function
procedure \cite{JSW} where relevant string surfaces ending at 
the AdS boundary saturate the correlators.
The extension to the three-point correlator of two heavy string vertex
operators with large charges and one light operator with fixed charges
has been presened in \cite{KZA}. Further constructions of the 
three-point correlators of two various heavy string states and a certain
light mode have been performed \cite{RH,AR,RHE}. This semiclassical
approach has been applied to the four-point correlators of two heavy
vertex operators and two light operators \cite{EBA} and also the
correlators involving Wilson loops and light local operators \cite{LAT}.

In the $\mathcal{N}=4$ SYM theory side the planar three-point correlators
of single trace gauge-invariant operators have been studied at weak
coupling \cite{OT}. Using the integrability techniques, the 
matchings of three-point correlators of two non-BPS operators and one
short BPS operator between weak and strong coupling have been 
demonstrated \cite{JE,GGE}.

The three-point correlator of heavy three BMN string states which are 
point-like in $AdS_5$ and rotating in $S^5$ has been computed by an 
assumption that the three cylinders associated with three external 
states may be joined together at an intersection point \cite{JSW}, where
the 4d conformal invariant dependence on the three positions of the
operator insertion points is extracted 
by taking advantage of the conformal
symmetry of $AdS_5$. Using a light-cone gauge for the worldsheet theory
the three-point correlator of BMN vertex operators has been constructed
by finding a solution in minimizing the action upon varying the 
intersection point of three Euclidean BMN strings \cite{KM}, where
 the three-point correlator of three heavy string vertex operators
for the circular winding strings with two equal spins in $S^5$ is also
computed to have a 4d conformal invariant expression.

Taking account of a string splitting process we have studied
the extremal three-point correlator of three heavy vertex operators for
the circular winding string states which are point-like in $AdS_5$
and rotating with two spins and two winding numbers in $S^5$ \cite{SR}.
A special correlator such that two of the three vertex operators are
located at the same point on the boundary 
 is evaluated on the stationary splitting string
surface consisting of three cylinders which is transformed by the
Schwarz-Christoffel map to the complex plane with three punctures which
are associated with the vertex insertion points.

There have been investigations of the AdS part of 
three-point correlator for the
string states without any spins in $AdS_2$ \cite{JW} as well as 
for the large spin limit of Gubser-Klebanov-Polyakov folded string states
rotating in  $AdS_3$ \cite{KK}, by using the Pohlmeyer reduction procedure
which is developed for the computation of gluon scattering amplitudes at
strong coupling. 

In ref. \cite{EA} the three-point correlators of three heavy vertex 
operators have been calculated for the ( BPS or non-BPS) string states
which have no spins in $AdS_5$ and carry  large charges in $S^5$
by using a particular Schwarz-Christoffel map defined by the AdS stress
tensor which transforms the three-geodesic solution in cylinder 
$(\tau,\sigma)$ domain to the complex plane with three punctures.
The resulting BPS correlator has been shown to agree with the large charge
limit of the corresponding supergravity and free gauge theory expressions.

We are interested in the three-point correlator for the heavy non-BPS 
string states which carry a spin in $AdS_5$. We will use a 
Schwarz-Christoffel map  to construct the extremal three-point
correlator of three heavy vertex operators for the circular spinning
strings carrying one large spin with one winding number in $AdS_5$
as well as one large spin with one winding number in $S^5$.
We will evaluate semiclassically the extremal three-point correlator
on a stationary string trajectory, where the three vertex operators are
located at the different points on the boundary. We will show that it 
becomes a 4d conformal invariant three-point correlator on the boundary.

Based on the vertex operator prescription \cite{AAT,EB} we consider an
extremal three-point correlation function of the three string 
vertex operators which are associated with the circular 
spinning string solutions with spins
and winding numbers $(S, n)$ in $AdS_5$ and $(J, m)$ in $S^5$. 

The embedding coordinates $Y_M \; (M=0, \cdots,5)$ of the Minkowski 
signature $AdS_5$ are expressed in terms of the global coordinates
$(t, \rho, \theta, \phi_1, \phi_2)$ as
\begin{eqnarray}
Y_5 + iY_0 &=& \cosh \rho e^{it}, \;\; Y_1 + iY_2 = \sinh \rho 
\cos \theta e^{i\phi_1}, \;\;Y_3 + iY_4 = \sinh \rho \sin \theta
e^{i\phi_2}, \nonumber \\
Y^MY_M &=& -Y_5^2 + Y^{\mu}Y_{\mu} + Y_4^2 = -1, \hspace{1cm} 
Y^{\mu}Y_{\mu} = -Y_0^2 + Y_mY_m
\label{em}\end{eqnarray}
with ${\mu}=0,1,2,3, m=1,2,3$. These coordinates are related with the
Poincare coordinates $(z, x^{\mu}), \;ds^2 = z^{-2}(dz^2 + 
dx^{\mu}dx_{\mu})$ as
$Y_{\mu} = x_{\mu}/z, \;\; Y_4 = (-1 + z^2 
+ x^{\mu}x_{\mu})/2z, \;\;\;\; Y_5 = (1 + z^2 + 
x^{\mu}x_{\mu})/2z$
with $x^{\mu}x_{\mu} = -x_0^2 + x_m^2$. For comparison we write down the
embedding coordinates of $S^5$,
$\rx_1 \equiv X_1 + iX_2 = \sin\gamma \cos\psi e^{i\varphi_1}, \;\;
\rx_2 \equiv X_3 + iX_4 = \sin\gamma \sin\psi e^{i\varphi_2}, \;\;
\rx_3 \equiv X_5 + iX_6 = \cos\gamma e^{i\varphi_3}$. 

Let us consider the vertex operator which describes the following
circular spinning closed string state with quantum numbers like spins 
$(S, J)$ and winding numbers $(n, m)$ \cite{ART}
\begin{eqnarray}
t &=& \kappa \tau, \;\; \rho = \rho_0, \;\; \theta = 0, \;\;
\phi_1 = \omega \tau + n \sigma \equiv \phi, \nonumber \\ 
\gamma &=& \frac{\pi}{2}, \;\; \psi = 0,  \;\; 
\varphi_1 = w\tau - m\sigma \equiv \varphi.
\label{sol}\end{eqnarray}
Its energy-spin relation is expressed in the $\la/J^2$ expansion as
\begin{equation}
E = J + S + \frac{\la}{2J^2}(m^2J + n^2S) 
- \frac{\la^2}{8J^5}( m^4J^2 + n^4SJ + 4n^2m^2SJ + 4n^4S^2) + \cdots.
\label{ejs}\end{equation}

Both the worldsheet time $\tau$ and the global AdS time $t$ 
are rotated to the Euclidean ones simultaneously, which lead to the 
similar rotations for the time-like coordinates $Y_0$ and $x_0$.
Using four coordinates $\vec{a}_i$ of a point on the
boundary of the Euclidean Poincare patch of $AdS_5$ space
we express the integrated vertex operator of dimension $\de_i$ as
\begin{equation}
V(\vec{a}_i) =  \int d^2\xi \left[ \frac{z}{z^2 + (\vec{x} - \vec{a}_i)^2}
\right]^{\de_i}(Y_1 + iY_2)^{S_i}\mathcal{V}^i_{AdS_5}
 (X_1 + iX_2)^{J_i}\mathcal{V}^i_{S^5},
\end{equation}
where $\vec{x}^2 = x_0^2 + x_m^2$ and $\mathcal{V}^i_{AdS_5},
 \;\; \mathcal{V}^i_{S^5}$
represent the derivative terms for the $AdS_5$ and $S^5$ parts that are 
not relevant for determining the stationary string trajectory.
We put the location of the vertex operator in the boundary 
as $\vec{a}_i = (a_i,0,0,0)$.
The expression $Y_1 + iY_2$ is described in terms of Euclidean Poincare
coordinates as 
$Y_1 + iY_2 = (x_1 + ix_2)/z = (r/z)e^{i\phi}$
and  the $S^5$ part is given by $X_1 + iX_2 = e^{i\varphi}$.
The winding number dependences are implicitly included through the
angular coordinates \cite{SR}.

We use the vertex operators labelled by points $a_1,a_2,a_3$ on the 
boundary of $AdS_5$ which are chosen as $a_1 = 0 < a_2 < a_3$
and compute a correlator of three heavy vertex operators
\begin{equation}
< V_{\de_1,S_1,n_1,J_1,m_1}(a_1) V^*_{\de_2,S_2,n_2,J_2,m_2}(a_2)
 V^*_{\de_3,S_3,n_3,J_3,m_3}(a_3) >
\label{cor}\end{equation}
in large spins of order of string tension $\sqrt{\la}/2\pi \gg 1$.
The Euclidean continuation of (\ref{sol}) is given by
$t = \kappa\tau, \;\; \rho = \rho_0,  \;\;
\phi = -i\omega\tau + n\sigma, \;\; \varphi = -iw\tau - m\sigma.$
The corresponding embedding coordinates in (\ref{em}) are expressed as
\begin{eqnarray}
Y_5 &=& \cosh\rho_0 \cosh\kappa\tau, \hspace{1cm} Y_0 =
 \cosh\rho_0 \sinh\kappa\tau, \hspace{1cm} Y_4 = 0, \nonumber \\
Y_1 &=& \sinh\rho_0 \cosh(\omega\tau + in\sigma), \hspace{1cm} 
Y_2 = -i\sinh\rho_0 \sinh(\omega\tau + in\sigma), \hspace{1cm} Y_3 = 0,
\end{eqnarray}
which are transformed to the Euclidean Poincare coordinates
in the form 
\begin{eqnarray}
z &=& \frac{1}{\cosh\rho_0 \cosh\kappa\tau}, \hspace{1cm}
x_0 = \tanh \kappa\tau, \nonumber \\
x_{\pm} &\equiv& x_1 \pm ix_2 = re^{\pm i\phi} = \frac{\tanh\rho_0}
{\cosh\kappa\tau} e^{\pm ( \omega \tau + in\sigma )}.
\label{tso}\end{eqnarray}
For the large spin limit $\kappa \approx \omega \gg 1$
 the complex world surface given by (\ref{tso}) approaches the boundary 
$z \rightarrow 0$ at $\tau \rightarrow \pm \infty$ with 
$x_0(\pm \infty) = \pm 1$ and $r(\pm \infty) = 0$.

In order to compute semiclassically the three-point correlator 
(\ref{cor}), we express the Euclidean action accompanied with the vertex
 contributions as an integral over the  complex $\xi$ plane 
\begin{eqnarray}
A &=& \frac{\sqrt{\la}}{\pi}\int d^2\xi \left[ \frac{1}{z^2}
( \pa z\bar{\pa}z + \pa x_0 \bar{\pa}x_0 + \pa r\bar{\pa}r
+ r^2\pa \phi\bar{\pa}\phi ) + \pa \varphi\bar{\pa}\varphi \right]
\nonumber \\
 &-& \int d^2\xi \left[ \sum_{i=1}^3 \de_i \del^2(\xi - \xi_i) 
\ln \frac{z}{z^2 + r^2 + (x_0 - a_i)^2}  \right] \nonumber \\
&-&  \int d^2\xi \left[ S_1\del^2(\xi - \xa) \ln \frac{re^{i\phi}}{z}
+ S_2\del^2(\xi - \xb) \ln \frac{re^{-i\phi}}{z} 
+ S_3\del^2(\xi - \xc) \ln \frac{re^{-i\phi}}{z} \right] \nonumber \\
&-&  \int d^2\xi [ J_1\del^2(\xi - \xa) \ln e^{i\varphi}
+ J_2\del^2(\xi - \xb) \ln e^{-i\varphi} 
+ J_3\del^2(\xi - \xc) \ln e^{-i\varphi} ],
\label{ac}\end{eqnarray}
where the $\ln \mathcal{V}_{AdS_5}^i$ and $\ln \mathcal{V}_{S^5}^i$ 
terms subleading in the
large spin are omitted and the vertex operators with dimensions 
$\de_1, \de_2, \de_3$ are located at the marked points $\xa, \xb, \xc$.

When the vertex operator with 
dimension $\de_1$ is inserted at $\tau =  -\infty$ and the other
two vertex operators with dimensions $\de_2$ and $\de_3$ are inserted
at $\tau =  \infty$ for the extremal case
\begin{equation}
\de_1 = \de_2 + \de_3,
\label{don}\end{equation}
the Schwarz-Christoffel map is described by
\begin{equation}
\zeta = e^{\tau + i\sigma} = \frac{\xi - \xa}
{(\xi - \xb)^{\de_2/\de_1}(\xi - \xc)^{\de_3/\de_1}}.
\label{sch}\end{equation}
The marked point $\xi_1$ is mapped to $\tau = -\infty$, while  $\xi_2$
and  $\xi_3$ are mapped to $\tau = \infty$.

In the $(\tau,\sigma)$ domain which is mapped to the upper half plane with
three marked points $\xa,\xb,\xc$ on the real axis by the 
Schwarz-Christoffel transformation (\ref{sch}) there are the following
three regions I, II, III which are identified with the three interacting
open strings and specified by the interaction points 
$(\tau_{int},\sigma_{int})$ in the open string picture
\begin{eqnarray}
I&:& -\infty < \tau \le \tau_{int}, \hspace{1cm} 0 \le \sigma \le \pi,
  \nonumber \\
II&:& \tau_{int} \le \tau < \infty, \hspace{1cm} 0 \le \sigma \le 
\sigma_{int} = \frac{\de_2}{\de_1} \pi, \nonumber \\
III&:& \tau_{int} \le \tau < \infty, \hspace{1cm} \sigma_{int}
  \le \sigma \le \pi,
\label{st}\end{eqnarray}
where $\tau_{int}$ is determined from the critical point of the map 
(\ref{sch}) $\pa \zeta /\pa \xi = 0$ as
\begin{equation}
\tau_{int} = \ln \frac{(\xb - \xa)^{\de_3/\de_1}(\xc - \xa)^{\de_2/\de_1}}
{\xc - \xb} + \ln \frac{\de_1}{\de_2^{\de_2/\de_1}\de_3^{\de_3/\de_1}}
\end{equation}
for $\xa < \xc < \xb$ \cite{EA}.
The closed string can be described by doubling trick, to take two copies
of the domain and perform appropriate identifications 
to ensure periodicity in $\sigma$.

For the string solution (\ref{tso})  we make the
respective dilatations and translations in the three regions
such that
\begin{eqnarray}
I&:&  x_0(\tau= -\infty) = a_1 = 0, \hspace{1cm} x_0(\tau= \infty) 
= b_1,  \nonumber \\
II&:&  x_0(\tau= -\infty) = b_2, \hspace{1cm} x_0(\tau= \infty) 
= a_2,  \nonumber \\
III&:&  x_0(\tau= -\infty) = b_3, \hspace{1cm} x_0(\tau= \infty) 
= a_3
\end{eqnarray}
at the boundary to obtain the independent solutions in the regions
I, II, III.
\begin{eqnarray}
I:  \;\; z &=& \frac{b_1}{2\cosh\rho_0 \cosh(\kappa_1\tau + \tau_1)},
 \hspace{1cm} x_0 = \frac{b_1}{2}(\tanh( \kappa_1\tau + \tau_1) + 1),
 \nonumber \\
x_{\pm} &=& re^{\pm i\phi} = \frac{b_1\tanh\rho_0}
{2\cosh(\kappa_1\tau + \tau_1)} e^{\pm ( \omega \tau + in\sigma )},
\nonumber \\
II:  \;\; z &=& \frac{a_2 - b_2}{2\cosh\rho_0 \cosh(\kappa_2\tau +
 \tau_2)},  \hspace{1cm} x_0 = \frac{a_2 - b_2}{2}\left(
\tanh( \kappa_2\tau + \tau_2) + \frac{a_2 + b_2}{a_2 - b_2} \right), 
 \nonumber \\
x_{\pm} &=& re^{\pm i\phi} = \frac{(a_2 - b_2)\tanh\rho_0}
{2\cosh(\kappa_2\tau + \tau_2)} e^{\pm ( \omega \tau + in\sigma )},
\nonumber \\
III:  \;\; z &=& \frac{a_3 - b_3}{2\cosh\rho_0 \cosh(\kappa_3\tau +
 \tau_3)},  \hspace{1cm} x_0 = \frac{a_3 - b_3}{2}\left(
\tanh( \kappa_3\tau + \tau_3) + \frac{a_3 + b_3}{a_3 - b_3} \right), 
 \nonumber \\
x_{\pm} &=& re^{\pm i\phi} = \frac{(a_3 - b_3)\tanh\rho_0}
{2\cosh(\kappa_3\tau + \tau_3)} e^{\pm ( \omega \tau + in\sigma )},  
\label{soz}\end{eqnarray}
where the parameters $\tau_i \; (i=1,2,3)$ are introduced.
The parameters $b_i, \; \tau_i$ will be discussed later to be
determined by the gluing condition at the three string interaction.
The Euclidean spinning string configuration in $S^5$ in each region 
is specified by
\begin{equation}
\rx_1 = e^{w\tau - im\sigma}.
\label{sso}\end{equation}

The angular momenta $S$ and $J$ for the closed string in each region are
evaluated by doubling the $\sigma$-integrals
\begin{eqnarray}
I&:& \;\; S_1 = 2\sqrt{\la}\int_0^{\pi} \frac{d\sigma}{2\pi} \omega 
\sinh^2\rho_0 = \sqrt{\la}\omega\sinh^2\rho_0, \;\;
J_1 =  2\sqrt{\la}\int_0^{\pi}\frac{d\sigma}{2\pi}w = \sqrt{\la}w, 
\nonumber \\
II&:& \;\; S_2 = 2\sqrt{\la}\int_0^{\sigma_{int}^{(1)}} 
\frac{d\sigma}{2\pi} \omega \sinh^2\rho_0 = \frac{\de_2}{\de_1}S_1,
\;\; J_2 =  2\sqrt{\la}\int_0^{\sigma_{int}^{(1)}} 
\frac{d\sigma}{2\pi}w = \frac{\de_2}{\de_1}J_1, \nonumber \\
III&:& \;\; S_3 = 2\sqrt{\la}\int_{\sigma_{int}^{(2)}}^{\pi}
 \frac{d\sigma}{2\pi} \omega \sinh^2\rho_0 = 
\frac{\de_3}{\de_1}S_1, \;\; J_3 =  2\sqrt{\la}
\int_{\sigma_{int}^{(2)}}^{\pi}\frac{d\sigma}{2\pi}
w = \frac{\de_3}{\de_1}J_1.
\label{sj}\end{eqnarray}
Owing to (\ref{don}) these charges are conserved 
\begin{equation}
S_1 = S_2 + S_3, \hspace{1cm} J_1 = J_2 + J_3.
\label{cco}\end{equation}
The continuity conditions of the string variables $Y_1 + i Y_2 =
\sinh \rho_0 e^{i\phi}$ and $\rx_1 = e^{i\varphi}$ at the
interaction time $\tau_{int}$ are satisfied in the same way
as discussed in \cite{SR}.

In the region I the parameter $m$ in (\ref{sso}) can be regarded as the
winding number of the incoming closed string $m_1 = m$.
By doubling the $\sigma$-intervals for the closed string configuration 
and rescaling the worldsheet space coordinate $\sigma$ to $\sigma_2, \; 
\sigma_3$ in such a way that $0\le \sigma \le 2\pi \de_2/\de_1 \rightarrow
0 \le \sigma_2 \le 2\pi$ and $2\pi \de_2/\de_1 \le \sigma \le 2\pi
\rightarrow 0 \le \sigma_3 \le 2\pi$ 
\begin{equation}
\sigma_2 = \frac{\de_1}{\de_2}\sigma, \hspace{1cm} 
\sigma_3 = \left(\sigma - 2\pi \frac{\de_2}{\de_1}\right)
\frac{\de_1}{\de_3},
\end{equation}
which are regarded as the worldsheet space coordinates of the regions
II, III we rewrite the expression (\ref{sso}) as the two outgoing 
string configurations in the regions II, III, 
$\rx_1^{II}(\sigma_2,\tau) = e^{w\tau - im_2\sigma_2}, \hspace{1cm}
\rx_1^{III}(\sigma_3,\tau) = e^{w\tau - im_3\sigma_3},$
where 
$m_2 =  (\de_2/\de_1)m, \;\; m_3 =  (\de_3/\de_1)m.$
The periodicity conditions for the outgoing closed string in the 
cylindrical regions II, III imply that $m_2, m_3$ are integers
and regarded as the respective winding numbers of the closed strings
in II, III. The winding numbers are also  conserved such that
$m_1 = m_2 + m_3$. In the winding number $n$ of the string configuration
in $AdS_5$ the same argument for $Y_1 + iY_2=\sinh\rho_0 e^{w\tau + 
in\sigma}$ yields the respective winding numbers in the three
cylindrical regions I, II, III
$n_1 = n, \;\; n_2 = (\de_2/\de_1)n, \;\; n_3 = (\de_3/\de_1)n.$

We will show that the relevant string configuration expressed in terms of
the complex worldsheet coordinate $\xi$  becomes the stationary string
trajectory in the presence of the vertex operators as source terms.
The equation of motion for $\phi$ is given by
\begin{equation}
\pa\left( \frac{r^2}{z^2} \bar{\pa}\phi \right) + 
\bar{\pa}\left( \frac{r^2}{z^2} \pa\phi \right) = - \frac{i\pi}
{\sqrt{\la}} [ S_1\del^2(\xi - \xa)  - S_2\del^2(\xi - \xb) - 
S_3\del^2(\xi - \xc)].
\label{eph}\end{equation}

The Schwarz-Christoffel map (\ref{sch}) is expressed as 
\begin{eqnarray}
\tau &=& \frac{1}{2}\left( \ln(\xi - \xa)(\bar{\xi} - \bar{\xa})
- \frac{\de_2}{\de_1} \ln (\xi - \xb)(\bar{\xi} - \bar{\xb})
- \frac{\de_3}{\de_1} \ln (\xi - \xc)(\bar{\xi} 
- \bar{\xc}) \right), 
\nonumber \\
\sigma &=&  \frac{1}{2i}\left( \ln \frac{\xi - \xa}
{\bar{\xi} - \bar{\xa}}
- \frac{\de_2}{\de_1} \ln \frac{\xi - \xb}{\bar{\xi} - \bar{\xb}}
- \frac{\de_3}{\de_1} \ln \frac{\xi - \xc}{\bar{\xi} - \bar{\xc}}
 \right),
\label{cs}\end{eqnarray}
which leads to
\begin{equation}
(\pa\bar{\pa} + \bar{\pa}\pa)\tau = \pi \left( \del^2(\xi - \xa)  - 
\frac{\de_2}{\de_1}\del^2(\xi - \xb)
- \frac{\de_3}{\de_1}\del^2(\xi - \xc) \right),  \hspace{1cm}
(\pa\bar{\pa} + \bar{\pa}\pa)\sigma = 0.
\label{tsp}\end{equation}
In the region I, $\xi$ cannot reach the points $\xb, \xc$ so that
$\del^2(\xi - \xb) = \del^2(\xi - \xc) =0$ and then the r.h.s. of
eq. (\ref{eph}) becomes $-i\pi S_1\del^2(\xi - \xa) /\sqrt{\la}$.
Substituting the region I string configuration of (\ref{soz}) and taking
account of (\ref{tsp}) we see that the l.h.s. of eq. (\ref{eph}) becomes
$-i\omega \sinh^2\rho_0\pi\del^2(\xi - \xa)$. Thus the eq.(\ref{eph})
in the region I is satisfied only when $S_1$ is given by   
\begin{equation}
S_1 = \omega \sinh^2\rho_0 \sqrt{\la}.
\label{sl}\end{equation}
Similarly the eq. (\ref{eph}) in the regions II, III also holds if the
respective spins $S_2, S_3$ are expressed as
$S_2 = (\de_2/\de_1)S_1, \;\; S_3 = (\de_3/\de_1)S_1,$
which are given in (\ref{sj}).

We consider the equation of motion for $x_0$ 
\begin{equation}
\pa\left( \frac{\bar{\pa}x_0}{z^2}\right) + \bar{\pa}\left( 
\frac{\pa x_0}{z^2} \right) = \frac{2\pi}{\sqrt{\la}} \sum_{i=1}^3
\de_i \frac{x_0 - a_i}{z^2 + r^2 + (x_0 - a_i)^2}\del^2(\xi - \xi_i)
\label{exa}\end{equation}
with $a_1 = 0$. This equation in the region I is satisfied when
\begin{equation}
\de_1 = \kappa_1 \cosh^2\rho_0 \sqrt{\la},
\label{dk}\end{equation}
while the equations in the regions II, III give 
$\kappa_2 \cosh^2\rho_0 \de_2/\de_1 = \de_2/\sqrt{\la},
\;\; \kappa_3 \cosh^2\rho_0 \de_3/\de_1 = \de_3/\sqrt{\la}.$
Thus we have
$\kappa_2 = \kappa_3 = \de_1/(\cosh^2\rho_0 \sqrt{\la})
= \kappa_1 \equiv \kappa.$

The equation of motion for $r$ reads 
\begin{eqnarray}
r\pa\left( \frac{\bar{\pa}r}{z^2}\right) + r\bar{\pa} \left(
\frac{\pa r}{z^2}\right) - 2\frac{r^2}{z^2}\pa \phi \bar{\pa}\phi &=& 
- \frac{\pi}{\sqrt{\la}}\sum_{i=1}^3 S_i \del^2(\xi - \xi_i)  \nonumber \\
&+& \frac{2\pi}{\sqrt{\la}}\sum_{i=1}^3 \de_i   \frac{r^2}{z^2 + r^2
 + (x_0 - a_i)^2}\del^2(\xi - \xi_i).
\label{re}\end{eqnarray}
In the region I, since $\tau \rightarrow -\infty$ corresponds to 
$\xi \rightarrow\xa$ the equation is given by
\begin{eqnarray}
- \kappa [ 2\kappa \pa\tau \bar{\pa}\tau + \tanh(\kappa\tau + \tau_1) 
(\pa\bar{\pa} + \bar{\pa}\pa)\tau ] - 2(n^2 - \omega^2)
\pa\tau \bar{\pa}\tau \nonumber \\
= \pi ( 2\kappa - \omega )\del^2(\xi - \xa),
\end{eqnarray}
where we use
$\pa\sigma \bar{\pa}\sigma = \pa\tau \bar{\pa}\tau, \;\;
\pa\tau \bar{\pa}\sigma + \pa\sigma \bar{\pa}\tau  = 0$
and (\ref{sl}), (\ref{dk}).  
The non-singular $\pa\tau \bar{\pa}\tau$ terms yield
\begin{equation}
\omega^2 = \kappa^2 + n^2
\label{omk}\end{equation}
and the remaining singular terms provide
\begin{equation}
-\kappa \tanh(\kappa \tau + \tau_1) \del^2(\xi - \xa) = 
 (2\kappa - \omega )\del^2(\xi - \xa) 
\label{kt}\end{equation}
through (\ref{tsp}).
Owing to  
$\tanh(\kappa \tau + \tau_1)|_{\xi \rightarrow \xa} = -1$
the eq. (\ref{kt}) is satisfied  to the leading order in the large
spin limit $\kappa \approx \omega \gg 1$.
In the regions II, III using the following limits
$\tanh(\kappa \tau + \tau_2)|_{\xi \rightarrow \xb} = 
\tanh(\kappa \tau + \tau_3)|_{\xi \rightarrow \xc} = 1,$
we see that the non-singular part gives the same relation as
(\ref{omk}) and  the singular part is also satisfied in the large
spin limit only when the relations between $S_i \; (i=1,2,3)$
are expressed as (\ref{sj}).

We turn to the equation of motion for $z$
\begin{eqnarray}
z\pa \left( \frac{\bar{\pa}z}{z^2}\right) + z\bar{\pa} \left( 
\frac{\pa{z}}{z^2}\right)  + \frac{2}{z^2}(\pa z \bar{\pa}z + 
\pa x_0 \bar{\pa}x_0 + \pa r \bar{\pa}r +
r^2\pa\phi \bar{\pa}\phi) \nonumber \\
=  \frac{\pi}{\sqrt{\la}} \sum_{i=i}^3 S_i\del^2(\xi - \xi_i) 
+ \frac{\pi}{\sqrt{\la}} \sum_{i=i}^3 \de_i \frac{z^2 - r^2 - 
(x_0 - a_i)^2}{z^2 + r^2 + (x_0 - a_i)^2}\del^2(\xi - \xi_i).
\label{ze}\end{eqnarray}
In the region I it becomes
\begin{eqnarray}
2\pa \tau\bar{\pa}\tau\sinh^2\rho_0 (\kappa^2 - \omega^2 + n^2)
- \kappa \tanh(\kappa \tau + \tau_1) 
(\pa\bar{\pa} + \bar{\pa}\pa)\tau \nonumber \\
= \pi[ \kappa( 1 - \sinh^2\rho_0 ) + \omega \sinh^2\rho_0]
\del^2(\xi - \xa),
\end{eqnarray}
which is satisfied in the large spin limit through
(\ref{omk}). In the regions II, III the equation 
(\ref{ze}) also holds owing to (\ref{omk}) and (\ref{sj}).

The remaining equation of motion for $\varphi$ is given by
\begin{equation}
(\pa\bar{\pa} + \bar{\pa}\pa)\varphi = -\frac{i\pi}{\sqrt{\la}}
[ J_1\del^2(\xi - \xa) - J_2\del^2(\xi - \xb) - J_3\del^2(\xi - \xc)],
\label{jee}\end{equation}
which combines with (\ref{tsp}) to yield the relations between 
$J_i \; (i=1,2,3)$ in (\ref{sj}).

Summing (\ref{re}) and (\ref{ze}) in order to eliminate 
the $\sum_i S_i\del^2(\xi - \xi_i)$ terms in the r.h.s. we have
\begin{eqnarray}
r\pa \left( \frac{\bar{\pa}r}{z^2}\right) + r\bar{\pa} \left( 
\frac{\pa r}{z^2}\right) +
z\pa \left( \frac{\bar{\pa}z}{z^2}\right) + z\bar{\pa} \left( 
\frac{\pa z}{z^2}\right)  + \frac{2}{z^2}(\pa z \bar{\pa}z + 
\pa x_0 \bar{\pa}x_0 + \pa r \bar{\pa}r) \nonumber \\
=  \frac{\pi}{\sqrt{\la}} \sum_{i=i}^3 \de_i \frac{z^2 + r^2 - 
(x_0 - a_i)^2}{z^2 + r^2 + (x_0 - a_i)^2}\del^2(\xi - \xi_i).
\label{co}\end{eqnarray}
We demand that the equation (\ref{exa}) and the combined equation 
(\ref{co}) are non-singular when $\xi \rightarrow \infty$.
The large $\xi$ behavior of the r.h.s. of (\ref{exa}) and (\ref{co})
suggests that for the coefficients of the common $\del^2(\xi)$
factor we should impose the two equations
\begin{eqnarray}
\de_1 \frac{x_0}{z^2 + r^2 + x_0^2}\bigg|_{\xi \rightarrow \xa} +
\de_2 \frac{x_0 - a_2}{z^2 + r^2 + (x_0 - a_2)^2}\bigg|_{\xi 
\rightarrow \xb} + \de_3 \frac{x_0 - a_3}{z^2 + r^2 + 
(x_0 - a_3)^2}\bigg|_{\xi \rightarrow \xc} = 0, \nonumber \\
\de_1 \frac{z^2 + r^2 - x_0^2}{z^2 + r^2 + x_0^2}\bigg|_{\xi 
\rightarrow \xa} + \de_2 \frac{z^2 + r^2 - (x_0 - a_2)^2}
{z^2 + r^2 + (x_0 - a_2)^2}\bigg|_{\xi \rightarrow \xb} +
\de_3 \frac{z^2 + r^2 - (x_0 - a_3)^2}{z^2 + r^2 + (x_0 - a_3)^2}
\bigg|_{\xi \rightarrow \xc} = 0.
\label{xz}\end{eqnarray}
These equations are indeed satisfied by the solution (\ref{soz}).

Here we require that the two equations in (\ref{xz}) hold at 
$\tau = \tau_{int}$ which is the common worldsheet time for the three
cylinders. Since $z^2 + r^2$ is expressed as $z^2\cosh^2\rho_0$
in each region, the interaction location $z(\tau_{int}) \equiv \tilde{z},
\; x_0(\tau_{int}) \equiv \tilde{x}_0$ is determined from the 
following equations
\begin{eqnarray}
\de_1 \frac{\tilde{x}_0}{\cosh^2\rho_0 \tilde{z}^2 + \tilde{x}_0^2} +
\de_2 \frac{\tilde{x}_0 - a_2}{\cosh^2\rho_0 \tilde{z}^2 + 
(\tilde{x}_0 - a_2)^2} + \de_3 \frac{\tilde{x}_0 - a_3}{\cosh^2\rho_0 
\tilde{z}^2 + (\tilde{x}_0 - a_3)^2} = 0, \nonumber \\
\de_1 \frac{\cosh^2\rho_0 \tilde{z}^2 - \tilde{x}_0^2}
{\cosh^2\rho_0 \tilde{z}^2 + x_0^2} + \de_2 \frac{\cosh^2\rho_0 
\tilde{z}^2 - (\tilde{x}_0 - a_2)^2}{\cosh^2\rho_0 \tilde{z}^2 + 
(\tilde{x}_0 - a_2)^2} + \de_3 \frac{\cosh^2\rho_0 \tilde{z}^2 - 
(\tilde{x}_0 - a_3)^2}{\cosh^2\rho_0 \tilde{z}^2 + (\tilde{x}_0 - a_3)^2}
= 0.
\label{dxz}\end{eqnarray}
This requirement is considered to be identical to treating the 
$\tau$-dependent string configuration within the subspace of
$(z, r, x_0)$ constrained by a relation $r = z\sinh\rho_0$ as
a point-like string. Up to a factor $\cosh\rho_0$ the equations
in (\ref{dxz}) take the same form as the equations derived by
extremizing the effective $AdS_5$ action for the three-point correlator
for the BPS states which is specified by the Witten diagram 
in supergravity description in ref. \cite{KM}, where the solution
of the equations is found. Following this solution we express
the relevant solution to the two equations in (\ref{dxz}) as
\begin{eqnarray}
\tilde{x}_0 &=& \frac{\al_1 a_2a_3(\al_2a_2 + \al_3a_3)}
{\al_1\al_2a_2^2 + \al_1\al_3a_3^2 + \al_2\al_3( a_3 - a_2 )^2},
\nonumber \\
\cosh\rho_0 \tilde{z} &=& \frac{\sqrt{\al_1\al_2\al_3( \al_1 + 
\al_2 + \al_3)}(a_3 - a_2 )a_2a_3 } { \al_1\al_2a_2^2 + 
\al_1\al_3a_3^2 + \al_2\al_3( a_3 - a_2 )^2},
\label{xa}\end{eqnarray}
where 
$\al_1 = \de_2 + \de_3 - \de_1, \;\;  \al_2 = \de_3 + \de_1 - \de_2, \;\;
\al_3 = \de_1 + \de_2 - \de_3.$
The two equations (\ref{eph}) and (\ref{jee}) have no
singularity at $\xi = \infty$ because of the charge
conservations in (\ref{cco}).

Now we shift $\tau$ to $\tau - \tau_{int}$ such that each $\tau$ in the
solution (\ref{soz}) for $AdS_5$ is replaced by $\tau - \tau_{int}$.
 Accordingly the solution (\ref{sso}) for $S^5$ is also expressed as
\begin{equation}
\rx_1 = e^{w(\tau - \tau_{int}) - im\sigma}.
\label{xs}\end{equation}
The shifted expressions of $z$ and $x_0$ in the three regions
are combined in such a way that the $\tau - \tau_{int}$ dependence is 
eliminated to yield
\begin{eqnarray}
I&:& \; \cosh^2\rho_0 z^2 = x_0(b_1 - x_0), \;\; II: \; 
\cosh^2\rho_0 z^2 = (a_2 - x_0)(x_0 - b_2), \nonumber \\
 III&:& \; \cosh^2\rho_0 z^2 = (a_3 - x_0)(x_0 - b_3).
\label{zxb}\end{eqnarray}
Demanding that at $\tau = \tau_{int}, (z, x_0)$ in (\ref{zxb}) become 
$(\tilde{z}, \tilde{x}_0)$ in (\ref{xa}) we determine 
$b_i\; (i=1,2,3)$ as
\begin{equation}
b_1 = \frac{( \al_2 + \al_3)a_2a_3}{\al_2a_2 + \al_3a_3}, \;\;
b_2 = \frac{ \al_1a_2a_3}{( \al_1 + \al_3)a_2 - \al_3a_3}, \;\;
b_3 = \frac{ \al_1a_2a_3}{( \al_1 + \al_2)a_3 - \al_2a_2}.
\end{equation}
Then the parameters $\tau_i\; (i=1,2,3)$ are fixed such that
each shifted expression of $(z, x_0)$ in the three regions I, II, III
(\ref{soz}) becomes $(\tilde{z}, \tilde{x}_0)$ at $\tau = \tau_{int}$
\begin{eqnarray}
\tau_1 &=& \frac{1}{2} \ln \frac{\al_1(\al_2a_2 + \al_3a_3)^2}
{( a_3 - a_2 )^2\al_2\al_3(\al_1 + \al_2 + \al_3)}, \hspace{1cm}
\tau_2 = \frac{1}{2} \ln \frac{a_3^2\al_1\al_3(\al_1 + \al_2 + \al_3)}
{\al_2(\al_3 a_3 - (\al_1 + \al_3) a_2)^2}, \nonumber \\
\tau_3 &=& \frac{1}{2} \ln \frac{a_2^2\al_1\al_2(\al_1 + \al_2 + \al_3)}
{\al_3(\al_2 a_2 - (\al_1 + \al_2) a_3)^2}.
\end{eqnarray}

The three-point correlator can be calculated semiclassically by evaluating
string action with source terms on the stationary string trajectory.
It is convenient to go back to the Euclidean cylindrical worldsheet 
coordinates $(\tau, \sigma)$ through the conformal transformation 
(\ref{sch}) for computing the string action in (\ref{ac})
\begin{eqnarray}
A_{str} &=& \frac{\sqrt{\la}}{4\pi} 2\left( \int_{\tau(\xa)}^{\tau_{int}}
d\tau \int_0^{\pi} d\sigma + \int_{\tau_{int}}^{\tau(\xb)}d\tau 
\int_0^{\sigma_{int}^{(1)}} d\sigma  +  \int_{\tau_{int}}^{\tau(\xc)}
d\tau \int_{\sigma_{int}^{(2)}}^{\pi} d\sigma 
\right)  L_{str}, \nonumber \\
L_{str} &=& \frac{1}{z^2} [ (\pa_{\tau}z)^2 + (\pa_{\tau}x_0)^2 + 
(\pa_{\tau}r)^2 + r^2( (\pa_{\tau}\phi)^2 + (\pa_{\sigma}\phi)^2 ) ] 
\nonumber \\
 &+&  (\pa_{\tau}\varphi)^2 + (\pa_{\sigma}\varphi)^2 
\label{as}\end{eqnarray}
with
$\tau(\xi_i) = ( \ln |\xi - \xa|^2 - \de_2/\de_1 \ln |\xi - \xb|^2
 - \de_3/\de_1 \ln |\xi - \xc|^2 )/2|_{\xi \rightarrow \xi_i}$,
where the integral region is divided into the three regions according 
to (\ref{st}). 

Substituting the shifted expression of each solution (\ref{soz}) in the
three regions I, II, III and (\ref{xs}) into the string action (\ref{as})
we derive 
\begin{eqnarray}
A_{str} &=& \frac{\sqrt{\la}}{2} \left[ -( \tau(\xa) - \tau_{int} ) + 
\frac{\de_2}{\de_1}( \tau(\xb) - \tau_{int} ) + 
\frac{\de_3}{\de_1}( \tau(\xc) - \tau_{int} ) \right] \nonumber \\
& \times & ( \kappa^2 \cosh^2\rho_0 + ( n^2 - \omega^2 )\sinh^2\rho_0
+ m^2 - w^2 ).
\end{eqnarray}
In this expression we should subtract the logarithmic divergences 
associated with self-contractions in the vertex operators.

The source terms in (\ref{ac}) are evaluated using the delta-function as
\begin{eqnarray}
A_{sour} &=& (\kappa\de_1 - \omega S_1 - wJ_1)( \tau(\xa) - \tau_{int}) 
- (\kappa\de_2 - \omega S_2 - wJ_2)( \tau(\xb) - \tau_{int}) \nonumber \\
&-& (\kappa\de_3 - \omega S_3 - wJ_3)( \tau(\xc) - \tau_{int}) 
+ \de_1\tau_1 - \de_2\tau_2 - \de_3\tau_3 \nonumber \\
&+& \de_1\ln b_1 + \de_2\ln (a_2 - b_2) + \de_3\ln (a_3 - b_3)\nonumber \\
&+& (\de_1 + \de_2 + \de_3) \ln \cosh\rho_0 - ( S_1 + S_2 +  S_3)
\ln \sinh\rho_0 \nonumber \\
&+& i[ -(nS_1 - mJ_1)\sigma(\xa) + (nS_2 - mJ_2)\sigma(\xb)
+ (nS_3 - mJ_3)\sigma(\xc) ]
\end{eqnarray}
with $\sigma(\xi_i) = \sigma|_{\xi \rightarrow \xi_i}$ in (\ref{cs}). 
Since the coefficients in the last two terms are expressed as
$nS_2 - mJ_2 = (nS_1 - mJ_1)\de_2/\de_1, 
nS_3 - mJ_3 = (nS_1 - mJ_1)\de_3/\de_1$,
we put 
\begin{equation}
nS_1 = mJ_1
\label{nsm}\end{equation}
to obtain the following three-point correlator that is consistent with
2d conformal symmetry
\begin{eqnarray}
&<& V_{\de_1,S_1,n_1,J_1,m_1}(a_1) V^*_{\de_2,S_2,n_2,J_2,m_2}(a_2)
 V^*_{\de_3,S_3,n_3,J_3,m_3}(a_3) > \nonumber \\
&\approx& \frac{(\sinh\rho_0)^{ S_1 + S_2 +  S_3} }
{(\cosh\rho_0)^{\de_1 + \de_2 + \de_3}}\frac{C_0}{a_2^{\al_3}a_3^{\al_2}
(a_3 - a_2)^{\al_1}} \int d^2\xa d^2\xb d^2\xc
f(|\xi_{12}|, |\xi_{23}|, |\xi_{13}|),
\label{vv}\end{eqnarray}
where 
\begin{eqnarray}
C_0 &=& \left( \frac{\al_1^{\al_1}\al_2^{\al_2}\al_3^{\al_3}
( \al_1 + \al_2 + \al_3 )^{ \al_1 + \al_2 + \al_3} }
{(\al_1 + \al_2)^{\al_1 + \al_2}(\al_1 + \al_3)^{\al_1 + \al_3}
(\al_2 + \al_3)^{\al_2 + \al_3} } \right)^{1/2}, \nonumber \\
f &=& \mathrm{exp} \biggl[ \frac{\sqrt{\la}}{2} \left(  
- \frac{\kappa \de_1}{ \sqrt{\la} }
 + \sinh^2\rho_0 (\omega^2 + n^2) + w^2 + m^2 \right)
\nonumber \\
&\times&  \left( \tau(\xa) - \tau_{int} - 
\frac{\de_2}{\de_1}(\tau(\xb) - \tau_{int}) - \frac{\de_3}{\de_1}
(\tau(\xc) - \tau_{int}) \right) \biggr].
\end{eqnarray}

Here we impose the marginality condition of vertex operator
\begin{equation}
- \frac{\kappa \de_1}{ \sqrt{\la}} + \sinh^2\rho_0 (\omega^2 + n^2) +
w^2 + m^2 = 0
\label{mm}\end{equation}
to observe that the leading contributions of the $\ln|\xi_i - \xi_j|$
terms vanish in the large spin. The three-point correlator (\ref{vv})
contains the 2-d conformal invariant subleading contribution
$|\xa - \xb|^{-2}|\xb - \xc|^{-2}|\xa - \xc|^{-2}$ coming from the
the derivative terms in the vertex operators, whose integration over
$\xa, \xb, \xc$ cancels against the Mobius group volume factor.
For (\ref{mm}) the elimination of $n$ and $\rho_0$ through (\ref{sl}),
(\ref{dk}), (\ref{omk}) leads to 
\begin{equation}
\frac{2\kappa\de_1}{\sqrt{\la}} - 2\omega \ms_1 - \kappa^2 = \mj_1^2 + m^2
\label{ma}\end{equation}
with $S_1 = \ms_1 \sqrt{\la}, \; J_1 = \mj_1 \sqrt{\la}$.
In ref. \cite{ART} the circular two-spin $(S,J)$ string solution with the
winding numbers $(n,m)$ was constructed to be specified
by the same relations between the relevant parameters as (\ref{sl}),
(\ref{dk}), (\ref{omk}) and $J_1 = w\sqrt{\la}$ in (\ref{sj}) 
where the off-diagonal Virasoro constraint gives the same relation as 
(\ref{nsm}) and the diagonal Virasoro constraint is presented by
\begin{equation}
\frac{2\kappa E}{\sqrt{\la}}  - 2\omega \ms - \kappa^2 = 
2\sqrt{m^2 + \nu^2} \mj - \nu^2 
\label{jn}\end{equation}
with $\mj = \sqrt{m^2 + \nu^2}$. 
We eliminate the parameter $\nu$ to see that 
the marginality condition (\ref{ma}) coincides with (\ref{jn}) when 
the dimension $\de_1$ is identified with the energy $E$ of the 
incoming string in the region I.
Therefore the dimension $\de_1$ is determined as (\ref{ejs}).

Thus we have the 4-d conformal invariant expression of
three-point correlator
\begin{equation}
\frac{C_3}{(a_2 - a_1)^{\de_1 + \de_2 - \de_3}
(a_3 - a_1)^{\de_1 + \de_3 - \de_2}(a_3 - a_2)^{\de_2 + \de_3 - \de_1} },
\end{equation}
where $a_1 = 0$ and the three-point coefficient is 
$C_3 = C_0 (\sinh\rho_0)^{ S_1 + S_2 +  S_3}/
(\cosh\rho_0)^{\de_1 + \de_2 + \de_3},$
where $C_0$ shows the same expression as the coefficients in the 
three-point correlator for the circular winding strings with two equal
spins in $S^5$ \cite{KM} as well as in the $AdS_5$ part of the
three-point correlator for the BPS strings with large three spins
in $S^5$ \cite{EA}. 
The normalized three-point coefficient is defined by using the two-point
coefficient $C_2(\de) = (\sinh\rho_0)^{2S}/(\cosh\rho_0)^{2\de}$ as
$\bar{C_3} = C_3/[C_2(\de_1)C_2(\de_2)C_2(\de_3)]^{1/2}$,
which becomes $C_0$.  

Taking account of the large $\mj$ expansion for $\kappa$
\begin{eqnarray}
\kappa &=& \mj + \frac{1}{2\mj} \left( m^2 + 2n^2 \frac{\ms}{\mj} \right)
\nonumber \\
&-& \frac{1}{8\mj^3} \left( m^4 + 4n^4\frac{\ms}{\mj} + 
8n^2m^2\frac{\ms}{\mj} + 12n^4\frac{\ms^2}{\mj^2} \right) + \cdots
\end{eqnarray}
we obtain the $\la/J^2$ expansions of the two factors
$\cosh^2\rho_0$ and $\sinh^2\rho_0$ which are 
contained in the three-point coefficient $C_3$ 
\begin{eqnarray}
\cosh^2\rho_0 &=& 
 1 + \frac{S}{J} - \frac{\la}{2J^2} \left( ( n^2 + m^2 )
\frac{S}{J} + 2n^2\frac{S^2}{J^2} \right)  \nonumber \\
&+& \frac{\la^2}{J^4} \left( \frac{3}{8}( n^2 + m^2 )^2\frac{S}{J} +
2n^2( n^2 + m^2 )\frac{S^2}{J^2} + \frac{5n^4}{2} \frac{S^3}{J^3}
\right) + \cdots,
\label{cos}\end{eqnarray}
which is regarded as the expression for the region I.
When $S$ is equal to $J$, the eq. (\ref{mm}) becomes
$\kappa^2 - \omega^2\sinh^2\rho_0 - n^2 = 0$,
which is expressed through (\ref{omk}) and (\ref{sl}) with $S_1 = S$ as
$S^2( 1 - \sinh^2\rho_0 ) = 2n^2 \la \sinh^4\rho_0$.
We obtain the compact expressions
\begin{equation}
\sinh^2\rho_0 = \frac{2}{1 + \sqrt{1 + 8\la\frac{n^2}{S^2}} }, \;\;
\cosh^2\rho_0 = \frac{3 + \sqrt{1 + 8\la\frac{n^2}{S^2}} }
{1 + \sqrt{1 + 8\la\frac{n^2}{S^2}} },
\end{equation}
that lead to the following energy-spin relation
\begin{equation}
E = \frac{3S + \sqrt{S^2 + 8\la n^2} }{2} \left( 
\frac{3S + \sqrt{S^2 + 8\la n^2} }{2(S + \sqrt{S^2 + 8\la n^2})} 
\right)^{1/2},
\end{equation}
whose $\la/S^2$ expansion reduces to (\ref{ejs}) with $S = J$.

Using the relations in (\ref{sj}) we rewrite
$\cosh^2\rho_0$ in (\ref{cos}) as
\begin{eqnarray}
\cosh^2\rho_0 &=& 1 + \frac{S_i}{J_i} - \frac{\la}{2J_i^2} \left( 
( n_i^2 + m_i^2 )\frac{S_i}{J_i} + 2n_i^2\frac{S_i^2}{J_i^2}
\right)  \nonumber \\
&+& \frac{\la^2}{J_i^4} \left( \frac{3}{8}
( n_i^2 + m_i^2 )^2\frac{S_i}{J_i} +
2n_i^2( n_i^2 + m_i^2 )\frac{S_i^2}{J_i^2} + \frac{5n_i^4}{2} 
\frac{S_i^3}{J_i^3}\right) + \cdots,
\end{eqnarray}
which takes the same value for $i = 1, 2, 3$, that is, each region of
three cylinders. In the same way from the eq. (\ref{ejs}) which is 
regarded as the incoming string energy in the region I, we express
the respective string energies in the three regions $i = 1, 2, 3$ 
in terms of the quantum numbers $S_i, n_i, J_i, m_i$ as
\begin{equation}
E_i = J_i + S_i + \frac{\la}{2J_i^2}(m_i^2J_i + n_i^2S_i) 
- \frac{\la^2}{8J_i^5}( m_i^4J_i^2 + n_i^4S_iJ_i + 4n_i^2m_i^2S_iJ_i 
+ 4n_i^4S_i^2) + \cdots.
\end{equation}

Using a Schwarz-Christoffel map from the complex plane with
three punctures to the string surface consisting of one cylinder and
two separated cylinders, we have computed an extremal
three-point correlator of heavy string vertex operators representing
the circular winding strings with spins $S, J$ and winding numbers 
$n, m$ in $AdS_5 \times S^5$.

We have performed three different transformations of scaling and
translation on the circular winding string solution in a cylinder
and constructed three string configurations in the three cylinders.
By gluing the three string configurations and making the
Schwarz-Christoffel transformation we have shown that the 
circular winding string configurations mapped on the complex 
plane with three punctures solve the relevant equations of motion on
the complex plane with the delta-function sources at the three 
insertion worldsheet points of vertex operators.

Combining the semiclassically evaluated string action with the vertex
contributions from the $AdS_5$ and $S^5$ parts we have observed
that the diagonal Virasoro constraint expression appears as a coefficient
of the worldsheet time interval between each vertex location time
$\tau(\xi_i), i = 1,2,3$ and the interaction time $\tau_{int}$, while
the off-diagnal Virasoro constraint expression appears as a 
coefficient of the worldsheet space coordinate 
$\sigma(\xi_i), i = 1,2,3$ at
each vertex operator. We have demonstrated that the marginality
condition of the vertex operator for the 2d scaling behavior of the
semiclassically evaluated three-point correlator yields the same
relation among energy (dimension), spins and winding numbers as
is obtained from the diagonal Virasoro constraint, while the requirement
of the three-point correlator to have 2d conformal invariant expression
gives the same relation between spins and winding numbers as follows
from the off-diagonal Virasoro constraint.

We have observed that the resulting extremal three-point correlator has 
the 4d conformal invariant dependence on the three different positions of 
vertex operators in the boundary. Although the three-point
coefficient is expressed explicitly by the dimensions 
of vertex operators, the spins and the winding numbers, 
the normalized three-point coefficient is described only by the
relevant dimensions which are specified by the  spins and  
the winding numbers.

\end{document}